\documentclass{article}
\usepackage{spconf,amsmath,graphicx}

\usepackage{graphicx}
\usepackage[T1]{fontenc}
\usepackage[utf8]{inputenc}

\usepackage{dsfont}
\usepackage{xcolor}

\usepackage{setspace}
\usepackage[colorlinks=true,linkcolor=blue,citecolor=red,hyperfootnotes,pdftex]{hyperref}
\usepackage{cite}

\usepackage{blindtext}

\usepackage{latexsym}
\usepackage{amssymb}

\usepackage{amsfonts}
\usepackage[mathscr]{eucal}
\usepackage{amsthm}

\usepackage{array}
\usepackage{multirow}

\usepackage{algorithmicx}
\usepackage{algorithm}
\usepackage{algpseudocode}

\usepackage{float}
\usepackage{url}
\usepackage{eurosym}
\usepackage{setspace}
\usepackage{balance}
\usepackage{dsfont}
\usepackage{bbm}
\usepackage{array}
\usepackage{caption}
\renewcommand\thetable{\arabic{table}}

\usepackage{multirow}

\def\ontop#1#2{\setbox0\hbox{#2}\copy0\llap{\raise\ht0\hbox{#1}}}

\usepackage{amsmath}

\DeclareFontFamily{U}{mathx}{\hyphenchar\font45}
\DeclareFontShape{U}{mathx}{m}{n}{<5> <6> <7> <8> <9> <10> <10.95> <12> <14.4> <17.28> <20.74> <24.88> mathx10}{}
\DeclareSymbolFont{mathx}{U}{mathx}{m}{n}
\DeclareMathSymbol{\bigtimes}{1}{mathx}{"91}

\DeclareMathAlphabet{\mts}{U}{rsfs}{m}{n}

\theoremstyle{definition}

\theoremstyle{plain}
\newtheorem{Thm}{Theorem}

\newtheorem{Asm}{Assumption}

\theoremstyle{remark}

\newcommand{\qedsymb}{\hfill\ensuremath{\blacksquare}}                 

\newcommand{\zbx}[1]{\raisebox{0pt}[0pt][0pt]{#1}}                     

\renewcommand{\,}{\hspace*{+1pt}}                                      

\usepackage{multirow}

\usepackage{booktabs} 

\definecolor{gr}{rgb}{0.50,0.50,0.50}
\definecolor{wh}{rgb}{1.00,1.00,1.00}

\algnewcommand{\Initial}              {\item[{\bf initialization:}]}
\algnewcommand{\Initiax}{\item[\textcolor{wh}{\bf initialization:}]}

\algrenewcommand{\algorithmicrequire}       {{\bf input:}}
\algnewcommand{\Requirx}{\item[\textcolor{wh}{\bf input:}]}

\algrenewcommand{\algorithmicensure}        {{\bf output:}}
\algnewcommand{\Outputx}{\item[\textcolor{wh}{\bf output:}]}

\algrenewcommand{\algorithmiccomment}[1]{\hfill\textcolor{gr}{$\triangleright$\ \zbx{#1}}}

\algrenewtext{If}[1]    {{\bf\algorithmicif\ }{#1}{\bf\ \algorithmicthen}}
\algrenewtext{ElsIf}[1] {{\bf\algorithmicelse\ \algorithmicif\ }{#1}{\bf\ \algorithmicthen}}
\algrenewtext{Else}     {{\bf\algorithmicelse}}
\algrenewtext{EndIf}    {{\bf\algorithmicend}}

\algrenewtext{For}[1]   {{\bf\algorithmicfor\ }{#1}{\bf\ \algorithmicdo}}
\algrenewtext{ForAll}[1]{{\bf\algorithmicfor\ all\ }{#1}{\bf\ \algorithmicdo}}
\algrenewtext{EndFor}   {{\bf\algorithmicend}}

\algrenewtext{While}[1] {{\bf\algorithmicwhile\ }{#1}{\bf\ \algorithmicdo}}
\algrenewtext{EndWhile} {{\bf\algorithmicend}}

\algrenewtext{Repeat}   {{\bf\algorithmicrepeat\ }}
\algrenewtext{Until}[1] {{\bf\algorithmicuntil\ }{#1}}

\algrenewcommand{\algorithmicindent}{8pt}

\usepackage{amsmath}
\title{Social Learning Under Inferential Attacks}
\name{Konstantinos Ntemos\qquad Virginia Bordignon\qquad Stefan Vlaski\qquad Ali H. Sayed\thanks{
                This work was supported in part by the Swiss National Science Foundation grant 205121-184999. E-mails: konstantinos.ntemos@epfl.ch, virginia.bordignon@epfl.ch, stefan.vlaski@epfl.ch,  ali.sayed@epfl.ch.}}

\address{School of Engineering, Ecole Polytechnique F\'ed\'erale de Lausanne (EPFL)}
\allowdisplaybreaks
\begin{document}
\ninept
\maketitle
\begin{abstract}
A common assumption in the social learning literature is that agents exchange information in an unselfish manner. In this work, we consider the scenario where a subset of agents aims at driving the network beliefs to the wrong hypothesis. The adversaries are unaware of the true hypothesis. However, they will ``blend in" by behaving similarly to the other agents and will manipulate the likelihood functions used in the belief update process to launch {\em inferential attacks}. We will characterize the conditions under which the network is misled. Then, we will explain that it is possible for such attacks to succeed by showing that strategies exist that can be adopted by the malicious agents for this purpose. We examine both situations in which the agents have minimal or no information about the network model.
\end{abstract}
%



\section{Introduction and related work}
In the {\em social learning} paradigm \cite{De_Groot_1974, Jadbabaie_2012, Zhao_2012, Molavi_2018,Lalitha_2014,Lalitha_2018,Nedic_2017,Salami_2017,Virginia_2020} agents aim at {\em learning} an underlying system state through their own observations as well as information provided by other agents. The communication among the agents is dictated by an underlying graph topology where each agent directly communicates only with its neighbors. The standard assumption in social learning literature is that agents exchange honestly their beliefs in a cooperative fashion. However, in many cases agents may exhibit intentional {\em misbehavior} or operate in a {\em faulty} manner. Many works studied the robustness of distributed processing systems against 
{\em Byzantine attacks} \cite{Lamport_1982, Strong_1982}, where adversaries can deviate from the system protocol in an {\em arbitrary} way. Another related line of research includes {\em robust detection} \cite{Huber_1965, Marano_2008, Kailkhura_2013, Vempaty_2013}, where adversaries aim at driving a fusion center to the wrong decision.

More recently, non-Bayesian social learning in the presence of malicious/faulty agents was studied in  \cite{Su_2018,Vyavahare_2019,Bhotto_2018,Hare_2019}. This problem poses new challenges, as each agent tries to learn the system state in a decentralized fashion. The authors in \cite{Su_2018, Vyavahare_2019} studied social learning in the presence of Byzantine attackers and showed that under certain assumptions on the structure of the communication graph, the normal agents successfully identify the underlying state. Other works studied more specific attack scenarios. One such scenario, which is the subject of interest in this paper, is driving the network beliefs to the wrong state. This cannot be guaranteed by letting adversaries send arbitrary information, since it can result in undesired non-convergent behavior. 
One pattern of adversarial behavior that may mislead the network is the use of {\em corrupted} likelihood functions when adversaries update their beliefs.

This kind of adversarial behavior was considered in \cite{Bhotto_2018,Hare_2019}. In \cite{Bhotto_2018} the authors devised a detection scheme where each agent reuses some of its most recent observations to update its beliefs along with the use of a convex combination of the beliefs of multiple social networks. The case where the agents' models are unknown and malicious agents are also present in the network was considered in \cite{Hare_2019}. The authors discuss different attack scenarios and discriminate between the so-called {\em weak} and {\em strong} malicious agents. Weak malicious agents manipulate only the likelihood functions in the belief update rule, while strong malicious agents can additionally filter out information sent from honest agents. The focus of these works is on detection of  adversaries. Instead, our work is focused on investigating adversarial strategies, which is not addressed by \cite{Bhotto_2018,Hare_2019}.

Our work addresses the case where the adversaries do not have knowledge about the true state and aim at forcing the network beliefs to the wrong hypothesis. We assume that adversaries participate in the information diffusion process as dictated by the social learning protocol, but disseminate falsified beliefs, which are produced by the use of corrupted likelihood functions. The fact that adversaries incorporate information from the network allows them to ``blend in" by appearing to follow the social learning protocol. We refer to this type of attacks as ``inferential attacks" due to the fact that adversaries have no knowledge of the true state and try to drive the network beliefs to the wrong state by manipulating their inference model (i.e, likelihood functions). We contribute to the related literature by investigating adversarial strategies.

More specifically, in this work, we answer the following questions. First, under what conditions is an unguarded network (no detection mechanism exists)  misled under inferential attacks? Second, if adversaries do not know the true state, is there a way to construct fake likelihood functions that drive the normal agents' beliefs to the wrong state? Finally, in scenarios of incomplete information, when adversaries do not have any knowledge about the network, how should they manipulate the observation models?

In the aforementioned setup, we characterize the conditions under which the network is misled. We prove that it depends on the agents' observation models, malicious and benign agents' centrality, and attack strategies. In this way, we reveal an interplay between the network topology, which captures the diffusion of information, and injection of mis-information in the social learning paradigm. Then, we prove that if an adversary knows certain network characteristics, then there is always an attack strategy that misleads the network. Finally, we study the scenario when adversaries have no knowledge about the network properties, propose an attack strategy, and illustrate its impact on the learning performance. 
\section{System Model}\noindent
We assume a set $\mathcal{N}=\mathcal{N}^n\bigcup\mathcal{N}^m$ of agents, where $\mathcal{N}^n$ and $\mathcal{N}^m$ denote the sets of normal and malicious agents, respectively. The types of the agents (i.e., normal or malicious) are unknown. The network is represented by an undirected graph $\mathcal{G}=\langle \mathcal{N}, \mathcal{E} \rangle$, where $\mathcal{E}$ includes bidirectional links between agents. The set of neighbors of an agent $k\in\mathcal{N}$ is denoted by  $\mathcal{N}_k$ (including agent $k$). 

We consider an adversarial setting where the normal agents aim at learning the true state $\theta^{\star}\in\Theta=\{\theta_1,\theta_2\}$, while malicious agents try to impede the normal agents by forcing their beliefs towards the wrong state. All agents are unaware of the true state $\theta^{\star}$.

We assume that each agent $k$ has access to observations $\boldsymbol{\zeta}_{k,i}\in\mathcal{Z}_k$ at every time $i\geq1$. Agent $k$ also has access to the likelihood functions $L_k(\zeta_{k,i}|\theta)$, $\theta\in\Theta$. The signals $\zeta_{k,i}$ are independent and identically distributed (i.i.d.) over time. The sets $\mathcal{Z}_k$ are assumed to be finite with $|\mathcal{Z}_k|\geq2$ for all $k\in\mathcal{N}$. We will use the notation $L_k(\theta)$ instead of $L_k(\boldsymbol{\zeta}_{k,i}\vert\theta)$ whenever it is clear from the context. 
\begin{Asm}{\bf (Finiteness of KL divergences)}. 
\label{obs_independence}
For any agent $k\in\mathcal{N}$ and for any $\theta\neq\theta^{\star}$,  $D_{KL}(L_k(
\theta^{\star})||L_k(
\theta))$ 
is finite.
\qedsymb\end{Asm}
At each time $i$, agent $k$ keeps a {\em belief vector} $\boldsymbol{\mu}_{k,i}$, which is a probability distribution over the possible system states. The belief components $\boldsymbol{\mu}_{k,i}(\theta)$, $\theta\in\Theta$, quantify the confidence of agent $k$ that each $\theta$ is the true state. We assume that all agents, both normal and malicious, are unaware of the true state. Thus, we impose the following assumption on initial beliefs.
\begin{Asm} {\bf (Positive initial beliefs)}. 
\label{positive_beliefs}
    $\mu_{k,0}(\theta)>0, \forall \theta\in\Theta,k\in\mathcal{N}$.
\qedsymb
\end{Asm}
\section{Social Learning with Adversaries}
Each normal agent $k$ uses the acquired observations $\boldsymbol{\zeta}_{k,i}$, along with the likelihood function $L_k(\zeta_{k, i}|\theta)$, to update their belief vector using Bayes' rule. Agents communicate with each other and exchange information. We consider the log-linear social learning rule \cite{Lalitha_2014,Virginia_2020,Matta_2020}. The normal agents update their beliefs in the following manner:
\begin{small}
\begin{align}
\label{adapt}
    &\boldsymbol{\psi}_{k,i}(\theta)=
    \frac{L_k(\boldsymbol{\zeta}_{k,i}|\theta)\boldsymbol{\mu}_{k,i-1}(\theta)}{\sum_{\theta'}L_k(\boldsymbol{\zeta}_{k,i}|\theta')\boldsymbol{\mu}_{k,i-1}(\theta')},\quad k\in\mathcal{N}^n\\
 \label{combine}
    &\boldsymbol{\mu}_{k,i}(\theta)=\frac{\prod_{\ell\in\mathcal{N}_k}\boldsymbol{\psi}^{a_{\ell k}}_{\ell,i}(\theta)}{\sum_{\theta'}\prod_{\ell\in\mathcal{N}_k}\boldsymbol{\psi}^{a_{\ell k}}_{\ell,i}(\theta')}, \quad k\in\mathcal{N}^n
\end{align}
\end{small}%
where $a_{\ell k}$ denotes the {\em combination weight} assigned by agent $k$ to neighboring agent $\ell$, satisfying $1\geq a_{\ell k}>0$, for all $\ell\in\mathcal{N}_k$, $a_{\ell k}=0$ for all $\ell\notin\mathcal{N}_k$ and $\sum_{\ell\in\mathcal{N}_k}a_{\ell k}=1$. Let $A$ denote the {\em combination matrix} which consists of all agents' combination weights. We impose the following assumption on the network topology.
\begin{Asm}{\bf{(Strongly-connected network)}.}
\label{strongly_connected}
The communication graph is {\em strongly connected} (i.e., there always exists a path with positive weights linking any two agents and at least one agent has a self-loop, meaning that there is at least one agent $k\in\mathcal{N}$ with $a_{kk}>0$).\qedsymb
\end{Asm}
For a strongly connected network, the limiting behavior of $A^{\mathsf{T}}$ is given by $\lim_{i\rightarrow\infty}(A^{\mathsf{T}})^i=\mathds{1}u^{\mathsf{T}}$, where $u$ is the Perron eigenvector {\color{blue}\cite{Sayed_2014}}. The eigenvector $u$ is associated with the eigenvalue at $1$, all its entries are positive and are normalized to add up to $1$. Moreover, its $k-$th entry $u_k$ expresses a measure of influence of agent $k$ on the network and it is also called the {\em centrality} of agent $k$.

We consider the scenario where adversaries aim at misleading the network to accept the wrong hypothesis by modifying the way they use their observations. More specifically, we assume that malicious agents deviate in step \eqref{adapt} by using a fake likelihood function, denoted by $\widehat{L}_k(\cdot)$ instead of $L_k(\cdot)$ to update their beliefs, while they follow \eqref{combine} 
without deviation. Inferential attacks are therefore modeled by assuming that adversaries follow the following update rule:
\begin{small}
\begin{align}
         \label{adapt_malicious}
    &\boldsymbol{\psi}_{k,i}(\theta)=
    \frac{\widehat{L}_k(\boldsymbol{\zeta}_{k,i}|\theta)\boldsymbol{\mu}_{k,i-1}(\theta)}{\sum_{\theta'}\widehat{L}_k(\boldsymbol{\zeta}_{k,i}|\theta')\boldsymbol{\mu}_{k,i-1}(\theta')},\quad k\in\mathcal{N}^m.
\end{align}
\end{small}%
We also impose the following technical assumption on the distorted likelihood functions.
\begin{Asm}
\label{finiteness}
\textbf{(Distorted likelihood functions with full support)}. For every agent $k\in\mathcal{N}^m$, the distorted likelihood function satisfies $\epsilon\leq\widehat{L}_k(\zeta_{k,i}|\theta)$ for all $\zeta_{k,i}\in\mathcal{Z}_k$, $\theta\in\Theta$, where $0<\epsilon\ll1$ is a small positive real constant that satisfies $\epsilon<\min_k\frac{1}{|\mathcal{Z}_k|}$.
\qedsymb\end{Asm}
We say that an agent $k$'s belief converges \textit{almost surely} (a.s.) to the true state if $\boldsymbol{\mu}_{k,i}(\theta^{\star})\to1$ as $i\to\infty$ with probability $1$. Conversely, agent $k$'s belief converges a.s. to the wrong state if $\boldsymbol{\mu}_{k,i}(\theta^{\star})\to0$ as $i\to\infty$ with probability $1$. The following result characterizes the asymptotic learning behavior of the network; the proof, as well as the proofs for subsequent results, are omitted due to space limitations.  
\begin{Thm}{\bf (Belief convergence with adversaries)}.
\label{inconsistent_learning}
Under Assumptions \ref{obs_independence}, \ref{positive_beliefs}, \ref{strongly_connected}, \ref{finiteness} the following are true:
\begin{enumerate}
    \item 
    The agents' beliefs converge a.s. to the wrong state if 
    \begin{small}
    \begin{align}
    \label{lem_ratiosm}
    \hspace{-8mm}\sum_{k\in\mathcal{N}^n}\hspace{-1.5mm}u_k\mathbb{E}\Bigg\{\log\frac{{L_k(\boldsymbol{\zeta}_k|\theta^{\star})}}{{L_k(\boldsymbol{\zeta}_k|\theta)}}\Bigg\}\hspace{-1mm}<\hspace{-1mm}
    \sum_{k\in\mathcal{N}^m}\hspace{-1.5mm}u_k\mathbb{E}\Bigg\{\log\frac{\widehat{L}_k(\boldsymbol{\zeta}_k|\theta)}{\widehat{L}_k(\boldsymbol{\zeta}_k|\theta^{\star})}\Bigg\}.
    \end{align}
        \end{small}
   \item The agents' beliefs converge a.s. to the true state if
    \begin{small}
    \begin{align}
    \label{lem_ratiosm2}
    \hspace{-8mm}\sum_{k\in\mathcal{N}^n}\hspace{-1.5mm}u_k\mathbb{E}\Big\{\log\frac{{L_k(\boldsymbol{\zeta}_k|\theta^{\star})}}{{L_k(\boldsymbol{\zeta}_k|\theta)}}\Big\}\hspace{-1mm}>\hspace{-1mm}
    \sum_{k\in\mathcal{N}^m}\hspace{-1.5mm}u_k\mathbb{E}\Big\{\log\frac{\widehat{L}_k(\boldsymbol{\zeta}_k|\theta)}{\widehat{L}_k(\boldsymbol{\zeta}_k|\theta^{\star})}\Big\}
    \end{align}
        \end{small}
\end{enumerate}
where $\theta^{\star},\theta\in\Theta$, $\theta^{\star}\neq\theta$.
\qedsymb
\end{Thm}
The Theorem characterizes under what condition the agents in the graph can be misled, namely, when condition \eqref{lem_ratiosm} holds. Thus, malicious agents would strive to construct their distorted likelihood functions to satisfy \eqref{lem_ratiosm}. The expectation in \eqref{lem_ratiosm} and \eqref{lem_ratiosm2} is taken w.r.t. the true likelihood distributions, $L_k(\boldsymbol{\zeta}_k|\theta^{\star})$. Since $\boldsymbol{\zeta}_{k,i}$ are i.i.d. over time, we omit the time index $i$. The threshold rule \eqref{lem_ratiosm}-\eqref{lem_ratiosm2} fully characterizes the convergence of network beliefs. Note that whether or not the agents' beliefs will converge to the true system state depends on agents' observation models (informativeness of the signals), on the distorted likelihood functions, and on  network topology (agents' centrality). 
\subsection{Attack strategies with known network divergence}
In this section, we answer the question of whether an adversary $k$ can construct $\widehat{L}_k(\theta_1),\widehat{L}_k(\theta_2)$ in such a way that the network will always be driven to the wrong hypothesis. Note that the state is unknown to adversaries. Thus, an adversary should select $\widehat{L}_k(\theta_1),\widehat{L}_k(\theta_2)$ such that \eqref{lem_ratiosm} is satisfied for both $\theta^{\star}=\theta_1$ and $\theta^{\star}=\theta_2$ to ensure that the network will converge to the wrong hypothesis no matter what the true state is. Let $S_j$ denote the term on the left-hand side of \eqref{lem_ratiosm} for $\theta^*=\theta_j, j=1,2$. Then, we can rewrite \eqref{lem_ratiosm} as follows:
    \begin{small}
    \begin{align}
    \label{lem_ratiosm30}
    &S_j
    <
    u_k\sum_{\zeta_k}L_k(\zeta_k|\theta_j)\log\frac{\widehat{L}_k(\zeta_k|\theta_{j'})}{\widehat{L}_k(\zeta_k|\theta_j)}
    , \quad k\in\mathcal{N}^m
\end{align}
    \end{small}\noindent
where $j,j'\in\{1,2\},j\neq j'$ and $\theta_j=\theta^{\star}$. We call $S_j$  \textit{network divergence}, or simply \textit{divergence}, of the normal subnetwork for $\theta^{\star}=\theta_j$.

Identifying a set of probability mass functions (PMFs) $\widehat{L}_k(\theta_1)$, $\widehat{L}_k(\theta_2)$ that mislead the network for both $\theta^{\star}=\theta_1$ and $\theta^{\star}=\theta_2$ requires solving the system of inequalities we get from \eqref{lem_ratiosm30} for $j=1,2$ w.r.t. $\widehat{L}_k(\zeta_k\vert\theta_1)$, $\widehat{L}(\zeta_k\vert\theta_2)$, $\zeta_k\in\mathcal{Z}_k$. In the following, we present one construction that captures such a family of PMFs. Before we present the result, we note that PMFs are \emph{uninformative} if  $L_k(\zeta_k\vert\theta_1)=L_k(\zeta_k\vert\theta_2)$ for all $\zeta_k\in\mathcal{Z}_k$, otherwise the PMFs are {\em informative}. Further, let us introduce the following quantities:
\begin{small}
\begin{align}
&n_j\triangleq L_{\ell}(\zeta^j_{\ell}\vert\theta_2)S_1+L_{\ell}(\zeta^j_{\ell}\vert\theta_1)S_2,\quad j=1,2\\
&d_{\ell}\triangleq L_{\ell}(\zeta^2_{\ell}\vert\theta_2)L_{\ell}(\zeta^1_{\ell}\vert\theta_1)-L_{\ell}(\zeta^2_{\ell}\vert\theta_1)L_{\ell}(\zeta^1_{\ell}\vert\theta_2).
\end{align}
\end{small}%
\begin{Thm}
\label{region}
    \textbf{(Distorted PMFs with known divergences)}. The following construction drives the network to the wrong hypothesis for any $\theta^{\star}\in\Theta$, given that there exists at least one adversary with informative PMFs, for sufficiently small $\epsilon$. Every adversary $\ell\in\mathcal{N}^m$ with informative PMFs uses the following construction.
\begin{small}
\begin{align}
\label{construction10}
    \widehat{L}_{\ell}(\zeta_{\ell}\vert\theta_j)=\begin{cases}\epsilon_{j'},\quad\,\text{ if } \zeta_{\ell}=\zeta^{j'}_{\ell},\\
    \alpha-\epsilon_{j'},\,\quad\text{ if } \zeta_{\ell}=\zeta^j_{\ell},\\
    \epsilon,\,\quad\text{ otherwise }
    \end{cases}
\end{align}
\end{small}%
where $\alpha=1-(|\mathcal{Z}_{\ell}|-2)\epsilon$, $j,j'\in\{1,2\}$, $j\neq j'$, $\zeta^1_{\ell},\zeta^2_{\ell}\in\mathcal{Z}_{\ell}$ are such that $L_{\ell}(\zeta^1_{\ell}\vert\theta_1)L_{\ell}(\zeta^2_{\ell}\vert\theta_2)\neq L_{\ell}(\zeta^1_{\ell}\vert\theta_2)L_{\ell}(\zeta^2_{\ell}\vert\theta_1)$ and 
\begin{small}
\begin{align}
\label{e1e2relation1}
        &\epsilon_1=\frac{e^{x_1}\alpha(e^{x_2}-1)}{e^{x_2}-e^{x_1}}\\
    \label{e1e2relation2}
    &\epsilon_2=\frac{\alpha(1-e^{x_1})}{e^{x_2}-e^{x_1}}
\end{align}
\end{small}%
where $x^+>x_1>x_{\ell,1}'$ if $d_{\ell}>0$ and $x^-<x_1<x_{\ell,1}'$ if $d_{\ell}<0$, $x^-=\log(\epsilon/(\alpha-\epsilon)$, $x^+=-x^-$, $x_{\ell,1}'=n_2/(u_{\ell}d_{\ell})$, $x_{\ell,2}'=n_1/(u_{\ell}d_{\ell})$  and
\begin{small}
\begin{align}
    x_2=\beta_{\ell} (x_1-x_{\ell,1}')+x_{\ell,2}',\quad |x_2|<x^+
\end{align}
\end{small}%
with $\beta_{\ell}$ such that
\begin{small}
\begin{align}
    \min_j\left\{-\frac{L_{\ell}(\zeta^1_{\ell}\vert\theta_j)}{L_{\ell}(\zeta^2_{\ell}\vert\theta_j)}\right\}<\beta_{\ell}<\max_j\left\{-\frac{L_{\ell}(\zeta^1_{\ell}\vert\theta_j)}{L_{\ell}(\zeta^2_{\ell}\vert\theta_j)}\right\}
\end{align}
\end{small}%
where $j=1,2$. If an adversary $\ell\in\mathcal{N}^m$ has uninformative PMFs, then it sets  $\widehat{L}_{\ell}(\theta_1)=\widehat{L}_{\ell}(\theta_2)=L_{\ell}(\theta_1)=L_{\ell}(\theta_2)$.
\qedsymb
\end{Thm}
\vspace{-0.1cm}
The above result states that even one adversary with informative likelihood functions can always construct fake PMFs that mislead the network, given that it has access to divergences $S_1,S_2$ of the normal subnetwork. 
\vspace{-0.15cm}
\subsection{Attack strategies with unknown network divergence}
In general, it is not realistic to assume knowledge of network divergences $S_1,S_2$ is available to the adversaries. Thus, in this section, we investigate what is the best that adversaries can do when they do not know the characteristics of the normal subnetwork. Rearranging \eqref{lem_ratiosm}, we can define the following cost function.
\begin{small}
\begin{align}
    \label{cost_function}
    &\mathcal{C}(\theta^{\star}
    )=\sum_{k\in\mathcal{N}^n}u_{k}D_{KL}(L_k(\theta^{\star}))||L_{k}(\theta))\nonumber\\
    &+\sum_{\ell\in\mathcal{N}^m}u_{\ell}\sum_{\zeta_{\ell}}L_{\ell}(\zeta_{\ell}|\theta^{\star})\log\frac{\widehat{L}_{\ell}(\zeta_{\ell}|\theta^{\star})}{\widehat{L}_{\ell}(\zeta_{\ell}|\theta)}
\end{align}
\end{small}\noindent
where $\theta^{\star},\theta\in\Theta, \theta_1\neq\theta_2$. We observe that the second term in \eqref{cost_function} is under malicious agents' control. Thus, one option for the adversaries is to minimize \eqref{cost_function} over $\widehat{L}_{\ell}(\theta_1),\widehat{L}_{\ell}(\theta_2)$. However, $\theta^{\star}$ is unknown as well. A viable alternative is to treat the true state $\theta^{\star}$ as a random variable, assuming some prior distribution over the states $\pi=(\pi_{\theta_1},\pi_{\theta_2})$. 
We assume that malicious agents share a common prior. Thus, taking expectation over the true state $\theta^{\star}$ in \eqref{cost_function} leads to the following minimization problem for the malicious agents:
\begin{small}
\begin{align}
    \label{opt_altao}
    &\min_{\widehat{L}_{\ell}(\theta_1),\widehat{L}_{\ell}(\theta_2)}\sum_{\theta\in\Theta}\pi_{\theta}C(\boldsymbol{\theta^{\star}}=\theta
    ),\,\,\ \ell\in\mathcal{N}^m
    \\
    &\text{s.t.}\,\,\, \widehat{L}_{\ell}(\zeta_{\ell}|\theta)\geq\epsilon,\quad\quad\quad\,\forall\zeta_{\ell}\in\mathcal{Z}_{\ell},\theta\in\Theta,
    \nonumber\\
    &\,\,\,\,\,\,\,\, \sum_{\zeta_{\ell}\in\mathcal{Z}_{\ell}}\widehat{L}_{\ell}(\zeta_{\ell}|\theta)=1,\quad\forall \theta\in\Theta.
    \nonumber
\end{align}
\end{small}%
It should be noted that the solution to the minimization problem above, denoted by  $\widehat{L}^{\star}_{\ell}(\theta_1),\widehat{L}^{\star}_{\ell}(\theta_2)$, minimizes the weighted average of \eqref{cost_function} for $\theta^{\star}=\theta_1$ and $\theta^{\star}=\theta_2$. This means that \eqref{lem_ratiosm} is not necessarily satisfied if adversaries utilize $\widehat{L}^{\star}_{\ell}(\theta_1),\widehat{L}^{\star}_{\ell}(\theta_2)$. We also highlight that the solution depends on the prior distribution of the true state $\pi$. The optimization problem is decomposable across agents $\ell\in\mathcal{N}^m$  
and thus \eqref{opt_altao} reduces to the following for each agent $\ell\in\mathcal{N}^m$:
\begin{small}
\begin{align}
\label{opt_prao2pr}
    &
    \hspace{-1mm}\min_{\widehat{L}_{\ell}(\theta_1)}\hspace{-1mm}\sum_{\zeta_{\ell}}\hspace{-0.5mm}Z_{\ell}(\zeta_{\ell})\hspace{-0.5mm}\log\widehat{L}_{\ell}(\zeta_{\ell}|\theta_1)\hspace{-1mm}-\hspace{-1.5mm}\max_{\widehat{L}_{\ell}(\theta_2)}\hspace{-1mm}\sum_{\zeta_{\ell}}\hspace{-0.5mm}Z_{\ell}(\zeta_{\ell})\hspace{-0.5mm}\log\widehat{L}_{\ell}(\zeta_{\ell}|\theta_2)\\
    &
    \text{s.t.}\,\,\,\,\,\,\, \widehat{L}_{\ell}(\zeta_{\ell}|\theta_1)\geq\epsilon,\,\,\,\,\,\quad\quad\quad
    \widehat{L}_{\ell}(\zeta_{\ell}|\theta_2)\geq\epsilon,\nonumber\\
    &\quad\,\,\,\sum_{\zeta_{\ell}\in\mathcal{Z}_{\ell}}\widehat{L}_{\ell}(\zeta_{\ell}|\theta_1)=1,\quad\sum_{\zeta_{\ell}\in\mathcal{Z}_{\ell}}\widehat{L}_{\ell}(\zeta_{\ell}|\theta_2)=1\nonumber
\end{align}
\end{small}\noindent
where
\begin{align}
Z_{\ell}(\zeta_{\ell})\triangleq\pi_{\theta_1}L_{\ell}(\zeta_{\ell}|\theta_1)-\pi_{\theta_2}L_{\ell}(\zeta_{\ell}|\theta_2),\, \quad\zeta_{\ell}\in\mathcal{Z}_{\ell}.
\end{align}
Note that each coefficient $Z_{\ell}(\zeta_{\ell})$ expresses the {\em relative confidence} that $\zeta_{\ell}$ resulted from state $\theta_1$ instead of $\theta_2$. We define the set
\begin{small}
\begin{align}
\mathcal{D}^1_{\ell}\triangleq\{\zeta_{\ell}:Z(\zeta_{\ell})
\geq0,\quad\ell\in\mathcal{N}^m\}
\end{align}
\end{small}\noindent
which is comprised of all observations $\zeta_{\ell}\in\mathcal{Z}_{\ell}$ for which the level of confidence that these observations were generated by state $\theta_1$ is greater than the confidence that they are generated by $\theta_2$. The set $\mathcal{D}^2_{\ell}=\mathcal{Z}_{\ell}\setminus\mathcal{D}^1_{\ell},\ell\in\mathcal{N}^m$ is comprised of the observations 
for which the confidence that they are generated from $\theta_2$ is greater compared to $\theta_1$. The solution depends on whether these sets are both non-empty or not. Next, we examine the two scenarios.
\vspace{-1.8mm}
\subsubsection{Mixed Confidence}
We study first the scenario where both sets $\mathcal{D}^1_{\ell},\mathcal{D}^2_{\ell}$ are non-empty, which means that some observations are more likely to have been generated by $\theta_1$, while some others by $\theta_2$. In this case, the solution to \eqref{opt_prao2pr} is given by the following result. 
\begin{Thm}
\label{opt_attackth}
\textbf{(Distorted PMFs with unknown divergences and mixed confidence)}. If both $\mathcal{D}^1_{\ell},\mathcal{D}^2_{\ell}$ are non-empty sets, then the attack strategy optimizing \eqref{opt_prao2pr} for an agent $\ell\in\mathcal{N}^m$ is given by
\begin{small}
\begin{align}
    \label{opt_attack}
    &\widehat{L}_{\ell}(\zeta_{\ell}|\theta_j)=\begin{cases} \epsilon,\quad\,\text{if } \zeta_{\ell}\in\mathcal{D}^j_{\ell},\\
                            \displaystyle
                    \frac{Z_{\ell}(\zeta_{\ell})(1-|\mathcal{D}^j_{\ell}|\epsilon)}{\sum\limits_{\zeta_{\ell}\notin\mathcal{D}^j_{\ell}}Z_{\ell}(\zeta_{\ell})},\quad\,\text{if } \zeta_{\ell}\notin\mathcal{D}^j_{\ell}
                            \end{cases}
\end{align}
\end{small}
where $j\in\{1,2\}$.%
\qedsymb\end{Thm}
The intuition behind the attack strategy is the following. We focus on the construction for $\widehat{L}_{\ell}(\theta_1)$ and the rationale is the same for $\widehat{L}_{\ell}(\theta_2)$. The constructed PMF is such that the least possible probability mass (i.e., $\epsilon$) is assigned to every observation $\zeta_{\ell}$ that is more likely to be generated from state $\theta_1$ (i.e., for all $\zeta_{\ell}\in\mathcal{D}^1_{\ell}$). 
For the remaining observations that are more likely to be generated from $\theta_2$ (i.e., $\zeta_{\ell}\in\mathcal{D}^2_{\ell}$) 
the probability mass placed on every $\zeta_{\ell}\in\mathcal{D}^2_{\ell}$ is in proportion to the difference in confidence that $\zeta_{\ell}$ is generated from $\theta_2$ instead of $\theta_1$. 
The more likely it is for $\zeta_{\ell}$ to be generated from $\theta_2$, the more probability mass is placed on $\widehat{L}(\zeta_{\ell}|\theta_1)$. 
\vspace{-1mm}
\subsubsection{Pure Confidence}
In the scenario where all observations are more likely to be generated from either $\theta_1$ or $\theta_2$, the solution is different, but the intuition remains similar. The solution to \eqref{opt_prao2pr}  
is the following.
\begin{Thm}
\label{opt_attack2}
\textbf{(Distorted PMFs with unknown divergences and pure confidence)}. 
Let $\mathcal{D}^1_{\ell}=\emptyset$ or $\mathcal{D}^2_{\ell}=\emptyset$. Then, the attack strategy optimizing \eqref{opt_prao2pr} for an agent $\ell\in\mathcal{N}^m$ is given by
\begin{small}
\begin{align}
    \label{opt_attacknuniform}
    &\widehat{L}_{\ell}(\zeta_{\ell}|\theta_j)=\begin{cases}1-(|\mathcal{Z}_{\ell}|-1)\epsilon,\,\quad\text{ if } \mathcal{D}^j_{\ell}=\mathcal{Z}_{\ell},\, 
    \zeta_{\ell}=\zeta_{min},\\ 
                                            \epsilon,\quad\,\text{ if } \mathcal{D}^j_{\ell}=\mathcal{Z}_{\ell}\,\text{ and } \zeta_{\ell}\neq\zeta_{min},\\
    \displaystyle\frac{Z_{\ell}(\zeta_{\ell})}{\sum\limits_{\zeta_{\ell}\in\mathcal{Z}_{\ell}}Z_{\ell}(\zeta_{\ell})},\quad\,\text{if } \mathcal{D}^j_{\ell}=\emptyset
                            \end{cases}
\end{align}
\end{small}\noindent
where $j\in\{1,2\}$ and 
$\zeta_{min}=\arg\min_{\zeta_{\ell}}\{Z_{\ell}(\zeta_{\ell})\}$.
\qedsymb
\end{Thm}
The intuition behind the result is the following. If $\mathcal{D}^2_{\ell}=\emptyset$, then it is more likely that all the observations are generated by state $\theta_1$. Thus, the PMF $\widehat{L}(\zeta_{\ell}|\theta_1)$ is generated according to the following rationale. The maximum possible probability mass is placed on the observation that is the least likely to be generated from state $\theta_1$, while the minimum possible probability mass $\epsilon$ is placed on the rest of the observations. Regarding the PMF $\widehat{L}_{\ell}(\zeta_{\ell}|\theta_2)$, the probability mass of every $\zeta_{\ell}$, is in proportion to the difference in confidence that $\zeta_{\ell}$ is generated from $\theta_1$ instead of $\theta_2$. The more likely $\zeta_{\ell}$ is to be generated from $\theta_1$ instead of $\theta_2$, the more probability mass is placed on $\widehat{L}_{\ell}(\zeta_{\ell}\vert\theta_2)$. The rationale is the same for the case $\mathcal{D}^1_{\ell}=\emptyset$.
\vspace{-1mm}
\section{Simulations}
\vspace{-1.2mm}
We assume $15$ agents, with $11$ normal and $4$ malicious agents, interacting over a strongly-connected network. Each agent assigns uniform combination weights to its neighbors. 
The agents observe the state through a binary symmetric channel (i.e., $\mathcal{Z}_k=\{\zeta_1,\zeta_2\}$ for all $k\in\mathcal{N}$) with observation probabilities $L_k(\zeta_1|\theta_1)=L_k(\zeta_2|\theta_2)=p$ and $L_k(\zeta_2|\theta_1)=L_k(\zeta_1|\theta_2)=1-p$. We set $\epsilon=10^{-3}$.
\begin{figure}[!h]
\centering
\includegraphics[width=0.4\textwidth]{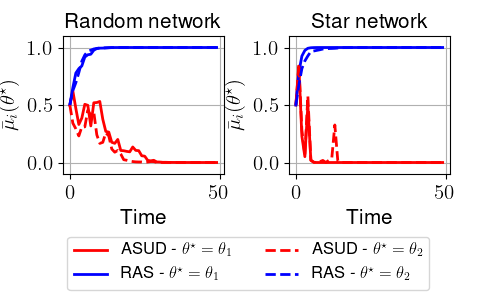}%
\vspace{-3mm}
\caption{Evolution of agents' average belief on $\theta^{\star}$ (i.e., $\bar{\boldsymbol{\mu}}_i(\theta^{\star})\triangleq\frac{\sum_{k\in\mathcal{N}}\boldsymbol{\mu}_{k,i}(\theta^{\star})}{|\mathcal{N}|}$) for $p=0.8$. Left: random topology, Right: star topology. ASUD: 
Attack Strategy with Unknown Divergences, RAS: Random Attack strategy.}
\label{network_ld}
\end{figure}
\begin{figure}[!h]
\centering
\includegraphics[width=0.4\textwidth]{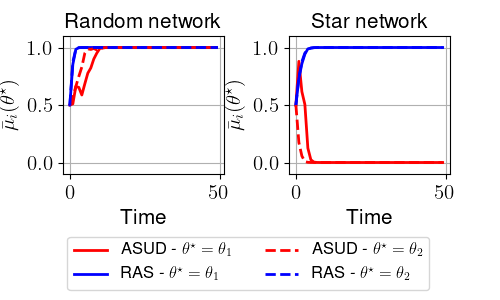}%
\vspace{-3mm}
\caption{Evolution of agents' average belief on $\theta^{\star}$ with highly discriminative models ($p=0.95$). Left: random topology, Right: star topology. 
}
\label{network_d}
\end{figure}
In our experiments we want to highlight the impact of adversarial strategies on the learning process in conjunction with the dependence on the  informativeness of agents' observation models and network topology. In Fig. \ref{network_ld} the binary symmetric channel is parametrized with $p=0.8$, while in Fig. \ref{network_d} agents' observation models are more discriminating between the two states with $p=0.95$. Apart from the dependence on the observation models, we demonstrate the impact of network topology by considering a random topology (left sub-figures) and star topology (right sub-figures). In the star topology, the central agent is malicious. In each case, we consider two different attack strategies, namely the attack strategy with unknown divergences presented in Theorem \ref{opt_attackth} for prior distribution $(\pi_{\theta_1},\pi_{\theta_2})=(0.5,0.5)$ and a random attack strategy, where the distortion functions $\widehat{L}(\theta_1),\widehat{L}(\theta_2)$ are chosen randomly by the malicious agents. As we observe in Fig. \ref{network_ld}, the network is misled under the proposed attack strategy for both network topologies in both cases when the system state is $\theta^{\star}=\theta_1$ and $\theta^{\star}=\theta_2$. The impact of random attack strategy is not sufficient to mislead the network.

The same rationale is followed in the experiments conducted for more discriminating models ($p=0.95$). As we observe in Fig. \ref{network_d}, the impact of malicious behavior is smaller in this setup, since normal agents are more capable to discriminate between the two hypotheses. More specifically, in the left sub-figure of Fig. \ref{network_d}, the network converges to the true state regardless of the attack type for the random network topology. On the other hand, for the star topology, where the central agent is malicious, the network is misled under the proposed attack strategy, as presented in the right sub-figure. This is because the overall centrality of the malicious agents is bigger in star topology compared to the random network topology. 
\vspace{-1mm}
\section{Conclusions}
\vspace{-1.5mm}
In this paper, the impact of inferential attacks on social learning was analyzed. We characterized the evolution of agents' beliefs and the adversaries' attack strategies were investigated. Our results are expected to shed light on the study of more elaborate attack schemes as well as on the development of light-weight detection mechanisms based on agents' characteristics (i.e., network centrality and observation models) and provide useful insight to situations where networks compete with each other in a strategic fashion.


\end{document}